\documentclass{nature}

\usepackage{amssymb}
\usepackage{graphicx}
\usepackage[FIGTOPCAP]{subfigure}
\usepackage{color}
\usepackage[normalem]{ulem}
\usepackage{soul}

\usepackage{subfig}

\definecolor{chinesered}{rgb}{0.828, 0.000, 0.000}

\usepackage{floatrow}
\DeclareFloatFont{tiny}{\scriptsize}
\title{A kilonova associated with GRB 070809}

\newcounter{firstbib}

\begin{document}

\maketitle
\author{Zhi-Ping Jin$^{1,2}$, Stefano Covino$^{3}$, Neng-Hui Liao$^{4,1}$, Xiang Li$^{1}$,  Paolo D'Avanzo$^{3}$, Yi-Zhong~Fan$^{1,2}$,  and Da-Ming Wei$^{1,2}$.}
\begin{affiliations}
\small
\item{Key Laboratory of Dark Matter and Space Astronomy, Purple Mountain Observatory, Chinese Academy of Sciences, Nanjing 210008, China}
\item{School of Astronomy and Space Science, University of Science and Technology of China, Hefei 230026, China}
\item{INAF/Brera Astronomical Observatory, via Bianchi 46, I-23807 Merate (LC), Italy}
\item{Department of Physics and Astronomy, College of Physics, Guizhou University, Guiyang 550025, China}
\end{affiliations}

\hfill

\begin{abstract}
For on-axis typical short gamma-ray bursts (sGRBs), the forward shock emission is usually so bright\cite{Eichler1989,Berger2014} that renders the identification of kilonovae (also known as macronovae)\cite{Li1998,Kasen2013,Tanaka2013,Metzger2017} in the early afterglow ($t<0.5$ d) phase rather challenging. This is why previously no thermal-like kilonova component has been identified at such early time\cite{Tanvir2013,Berger2013,Yang2015,Jin2015,Jin2016,Troja2018,Jin2018} except in the off-axis dim GRB 170817A\cite{Goldstein2017,Pian2017,Drout2017,Kasliwal2017,Arcavi2017,Cowperthwaite2017} associated with GW170817\cite{Abbott2017}. Here we report the identification of an unusual optical radiation component in GRB 070809 at $t\sim 0.47$ d, thanks plausibly to the very-weak/subdominant forward shock emission. The optical emission with a very red spectrum is well in excess of the extrapolation of the X-ray emission that is distinguished by an unusually hard spectrum, which is at odds with the forward shock afterglow prediction but can be naturally interpreted as a kilonova. Our finding supports the speculation that kilonovae are ubiquitous\cite{Jin2016}, and demonstrates the possibility of revealing the neutron star merger origin with the early afterglow data of some typical sGRBs that take place well beyond the sensitive radius of the advanced gravitational wave detectors\cite{Abadie2010,LiX2016} and hence the opportunity of organizing dedicated follow-up observations for events of interest.
\end{abstract}
%


To monitor the distant kilonovae well, clues emerging at early time are essential to organize the worldwide dedicated observations
since usually the late optical emission are so dim
that it requires the largest size ground-based instruments or the Hubble Space Telescope (HST).
The challenge is that at such early times usually the forward shock optical afterglow emission of the on-axis GRBs is usually so bright that it would outshine the kilonova component. In this work we show that such a goal has been achieved for the first time in GRB 070809, a typical sGRB with promptly detected X-ray afterglow that is expected to be an on-axis event. The kilonova signal was identified at $t\sim 0.47$ d (throughout this work t? represents the time measured in the source frame), much earlier than all previous kilonova candidates\cite{Tanvir2013,Berger2013,Yang2015,Jin2015,Jin2016,Jin2018,Troja2018}.
Moreover, GRB 070809 plausibly took place outside of its host galaxy and there is no convincing evidence for a regular forward shock emission of either the X-ray afterglow or the optical afterglow, , in contrast to other on-axis events.

GRB 070809 was discovered with the Burst Alert Telescope (BAT) on board the {\it Swift} satellite and the follow-up X-ray and optical observations were carried out with a number of telescopes\cite{Marshall2007,Perley2007a,Perley2007,Berger2010}.  We have downloaded and analysed the public archive data of Keck, Gemini, HST and X-ray Telescope (XRT).  The details of the data analysis are described in the Methods. Generally speaking, our results (see Supplementary Table 1) are nicely in good agreement with those reported in the literature \cite{Marshall2007,Perley2007a,Berger2010}.
With the HST data, we rule out the presence of a host galaxy coincident with the transient location down to 28.0 mag (AB) in the F606W band (see Fig.1 and Supplementary Table 1). For GRB 070809 there are two host galaxy candidates. One is an edge-on spiral galaxy centred at an offset of $5.9''$ to the northwest of GRB 070809 that has a redshift of $z=0.2187$ (ref.\cite{Perley2007}). The other is an early type galaxy at an offset of $6.0''$ that has $z=0.473$ (ref.\cite{Berger2010}). The latter has a lower possibility ($\approx 2.5\%$) of chance coincidence than the former ($\approx 10\%$) but the difference is just by a factor of $\sim 4$ (ref.\cite{Berger2010}), indicating that the redshift is not secure. We therefore adopt both in the discussion. At $z=0.2187~(0.473)$, GRB 070809 would have an isotropic-equivalent gamma-ray radiation energy $E_\gamma>10^{49}~(5\times 10^{49})$ erg, respectively (we set a lower limit because of narrow energy range of {\it Swift} BAT), ranking as a ``typical"/luminous short event.

The multiband afterglow light curves and spectral energy distribution (SED) of GRB 070809 are presented in Fig.2. In this work we adopt the $\Lambda$ cold dark matter model cosmology (where $\Lambda$  is the cosmological constant) with parameters of Hubble constant $H_{0}=67.4\pm0.5 {\rm km}~{\rm s}^{-1}~{\rm Mpc}^{-1}$ and the matter density $\Omega_{\rm m}= 0.315\pm0.007$ (ref.\cite{Planck2018}). In comparison with the X-ray emission of GW170817/GRB170817A emerging at $t>10$ d, GRB 070809 had an early X-ray afterglow starting at $t\leq 71$ s. Such a difference is most probably due to the off-axis nature of GW170817/GRB 170817A (for discussion of the outflow structure of this event please see some recent references\cite{Alexander2018,Piro2019}). Intriguingly, the light curves of  the optical emission following GW170817/GRB 170817A (that is, AT2017gfo) and GRB 070809 may be rather similar (see Fig.2a for the particular case of $z=0.2187$). The optical to X-ray SED of GRB 070809 measured at $t\sim 0.47$ d is distinguished by a rather red optical component (that is, $f_{\nu_{\rm opt}}\propto \nu^{-2.52\pm0.86}$, hereafter $\nu$ is the observed frequency and $\nu '=(1+z)\nu$ is the intrinsic frequency) and a very hard X-ray component (i.e., $f_{\nu_{\rm X}}\propto \nu^{-0.22\pm0.20}$). In Fig.3 we summarize the optical to X-ray SED of some bright/typical sGRBs with relatively plentiful afterglow data, for which a reliable SED at $t'<1$ d can be constructed; see Methods for the details. Evidently, GRB 070809 has the hardest X-ray spectrum and the softest optical spectrum. The two components, moreover, cannot be extrapolated to obtain a typical synchrotron spectrum, as the optical component is much brighter than any X-ray extrapolation. This is remarkably different from the case of GRB 130603B, for which the observed optical spectrum is very red due to the serious  extinction of the host galaxy but the X-ray spectral extrapolation to the optical bands well exceeds the observed ones (see Fig.5 of ref.\cite{deUgartePostigo2014}).
In Fig.3 we find similarities among GW170817/AT2017gfo, GRB 070809 (particular if $z=0.2187$), GRB 160821B and potentially also GRB 061201, which might be evidence for a similar physical origin and indicate that kilonovae are ubiquitous. Please note that for both GRB 061201 and GRB 160821B, the optical to X-ray SEDs at $t\leq 0.1$ d are well described as single power laws and in agreement with the forward shock afterglow model. If such SEDs hold for the latter emission as well, there are distinct near-infrared/optical forward shock radiation excesses which may be interpreted as the kilonova signals (see Supplementary Fig. 1 and  Methods for the details).

As extensively investigated in the Methods, the optical and X-ray emission of GRB 070809 should have different physical origins. The optical spectrum is unusually red. The introduction of a high extinction of the ``host" galaxy would result in a hard intrinsic optical spectrum. A high extinction in the host galaxy, however, is not supported by the very low intrinsic hydrogen column density of $N_{\rm H}=0^{+3.42}_{-0}\times10^{20}{\rm cm}^{-2}$. Moreover, all short/long-short bursts suffering from strong extinction were identified within or on the outskirts of the bright host galaxies (see Supplementary Table 2). For GRB 070809, no host galaxy has been identified down to $28$th mag. At a redshift of $z=0.2187$ or $0.473$, such a stringent constraint can only be met by a faint dwarf galaxy like those belonging to the Local Group or if the burst took place outside the host galaxy, no strong extinction is expected in either case.  Therefore, the forward shock afterglow interpretation, in the absence of motivation for assuming strong extinction, requires an extremely large power-law index of the accelerated electrons (i.e., $p\sim 5$), which in turn would yield a decline of $\propto t^{-3}$, too quick to be consistent with the current optical data (note that the $t\sim 0.001-0.5$ d $R-$band data/upper limits require a decline shallower than $t^{-0.96}$). Instead, the optical emission can be naturally interpreted as a kilonova at $z=0.2187~(0.473)$ with intrinsic temperature $T_{\rm int}\sim 5380~(7040)$K and bolometric luminosity $L\approx 0.63~(3.8)\times 10^{42}~{\rm erg~s^{-1}}$.
The bulk Lorentz factor $\Gamma$ (the corresponding velocity, in unit of the speed of light ($c$), is denoted as $\beta_{\Gamma}$) of the outflow is governed by  $(1+\beta_{\Gamma})\beta_{\Gamma}\Gamma \approx 0.9L_{42}^{1/2}(T_{\rm int}/6000~{\rm K})^{-2}(1+z)(t/0.47~{\rm d})^{-1}$. Supposing the observation time is comparable to the diffusion $(1+z)(\kappa M_{\rm ej}/4\pi \beta_{\Gamma}c^{2})^{1/2}$,  the mass of the wind, can be estimated as $M_{\rm ej}\sim 0.018~{M_\odot}~(\beta_{\Gamma}/0.5)(t/0.5~{\rm d})^{2}[(1+z)/1.22]^{-2}(\kappa/0.2~{\rm cm^2~g^{-1}})^{-1}$, where $\kappa$ is the opacity parameter of the sub-relativistic lanthanide-poor/free outflow.
For $z=(0.2187,~0.473)$, we have $\beta_{\Gamma}\approx (0.51^{+0.07}_{-0.15},~0.72^{+0.09}_{-0.2})$ and $M_{\rm ej}\sim 0.018~(0.016)M_\odot$.
Evidently, if $z=0.2187$ the kilonova signal of GRB 070809 well resembles the early blue component of AT2017gfo\cite{Pian2017,Drout2017,Kasliwal2017,Arcavi2017,Cowperthwaite2017} and may be powered by the lanthanide-poor disk wind, while for $z=0.473$, the yielded $\beta_\Gamma$ is so high that it probably favors the shock breakout/cocoon model\cite{Kasliwal2017}.
The X-ray emission, characterized by a unusual hard spectrum and the complicated temporal behavior, probably originates from the prolonged magnetic activity of the central engine.
It is surprising to notice that neither the X-ray nor the optical emission can be reasonably interpreted as the regular forward shock afterglow, because all other GRBs with kilonova signals (except the off-axis event GRB 170817A/AT2017gfo) have significant forward shock afterglow emission in X-ray and/or optical bands (see Fig.3). In any case, the absence of forward shock emission is well anticipated if a merger event takes place outside the host galaxy. As found in population synthesis studies, binary neutron star systems have a typical kick velocity of $v\approx 100~{\rm km~s^{-1}}$. For a merger timescale of $\sim 0.1-1$ Gyr, an offset $\sim 10-100$ kpc of the sGRB/kilonova from the host galaxy is expected. Therefore, it would be quite natural to have some GRBs/kilonovae outside the host galaxies. Previously some hostless sGRBs have been reported but it is unclear whether these events are indeed outside their host galaxies (see Berger\cite{Berger2010} for a through examination). The absence of a underlying host galaxy down to very stringent limits and the lack of clear evidence for forward shock emission in the afterglow data likely render GRB 070809 the best candidate for the population of  "outside" events.

Previously,  ($T_{\rm int}$) values have been reported in GRB 060614\cite{Jin2015}, GW170817/AT2017gfo\cite{Drout2017} and tentatively GRB 160821B\cite{Jin2018}, from which the luminosities ($L$) and the velocities ($\beta_\Gamma$) can be reasonably inferred, while for GRB 130603B and GRB 050709, we can only set upper limits on $T_{\rm int}$. Fig.4 summarizes the properties of the current kilonova event and candidates (see Methods for the details). We find interesting similarities between GW170817/AT2017gfo and the kilonova signal of GRB 160821B in the evolutions of $L,~T_{\rm int}$ and $\beta_\Gamma$. The kilonova signals in GRB 070809 and possibly also GRB 061201 displayed a little earlier than the start of the observation of AT2017fgo. The $T_{\rm int}$ ($\beta_\Gamma$) of these two signals are probably lower (higher) than that of AT2017gfo at similar epochs, indicating the diversity and potentially also the diverse origins of the kilonova emission (components).

The kilonova sample increases continually and at early times important emission are in optical bands (see Fig.4),
which provide the valuable chance to reveal the neutron star merger origin of some GRBs that are well beyond the sensitive distance of  advanced gravitational wave detectors such as LIGO and Virgo. This could be the case in particular for some distant sGRBs with very faint forward shock optical afterglow emission. Quick follow-up photometric observations of such sGRBs could yield significant evidence for the emergence of a kilonova at earlier times, as reported in this work for GRB 070809. For similar intriguing events discovered in the future, worldwide follow-up observations with large size ground-based telescopes can be effectively organized and the late time HST observations should be carried out to catch the red kilonova component. Most of the merger-driven GRBs with GW data from advanced LIGO/Virgo will appear as underluminous events since the energetic outflow core will be viewed off-axis, while the typical sGRBs at relatively high redshifts are usually on-axis events. A large GRB/kilonova sample consisting of both types of event would be essential in examining the angular dependence of the ejecta (including the nucleosynthesis products) and the generality of such a kind of association.



\renewcommand{\refname}{References}


\begin{addendum}

\item[Acknowledgements] This work was supported in part by NSFC under grants of No. 11525313 (i.e., Funds for Distinguished Young Scholars), No. 11433009 and No. 11773078, the Funds for Distinguished Young Scholars of Jiangsu Province (No. BK20180050), the Chinese Academy of Sciences via the Strategic Priority Research Program (Grant No. XDB23040000), Key Research Program of Frontier Sciences (No. QYZDJ-SSW-SYS024). SC and PD have been supported by ASI grant I/004/11/0.

\item[Author Contributions] Y.Z.F, Z.P.J, S.C, and D.M.W launched the project. Z.P.J, N.H.L, X.L (from PMO), S.C and P.D (from INAF/OAB) carried out the data analysis. Y.Z.F and D.M.W interpreted the data. Y.Z.F and Z.P.J prepared the paper and all authors joined the discussion.


\item[Author Information] Reprints and permissions information is available at www.nature.com/reprints. The authors declare no competing financial interests.
Correspondence and requests for materials should be addressed to Y.Z.F (yzfan@pmo.ac.cn).

\newpage

\begin{figure}[!h]
\begin{center}
\includegraphics[width=1.0\textwidth]{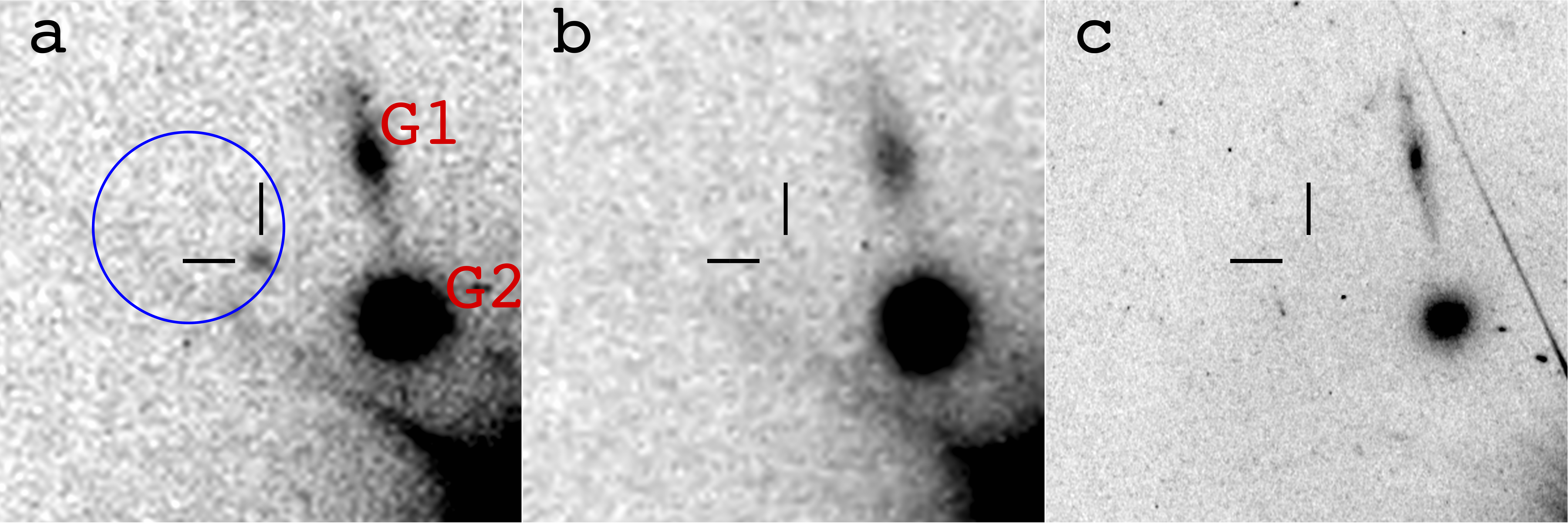}
\end{center}
\caption{{\bf Keck and HST observations of GRB 070809.} {\bf a}, Keck observation in the $R$-band at 0.47 d after the burst. The optical afterglow position is marked with red lines, and the X-ray error box is circled in blue. G1 and G2 are the possible host galaxies at $z=0.2187$ and $z=0.473$, respectively. {\bf b}, Keck observation in $R$-band at 1.47 d after the burst. {\bf c}, HST observation in $F606W$-band at 731 d after the burst.
}
\label{fig:GRB070809field}
\end{figure}

\newpage

\begin{figure}[!h]
\begin{center}
\includegraphics[width=1.0\columnwidth]{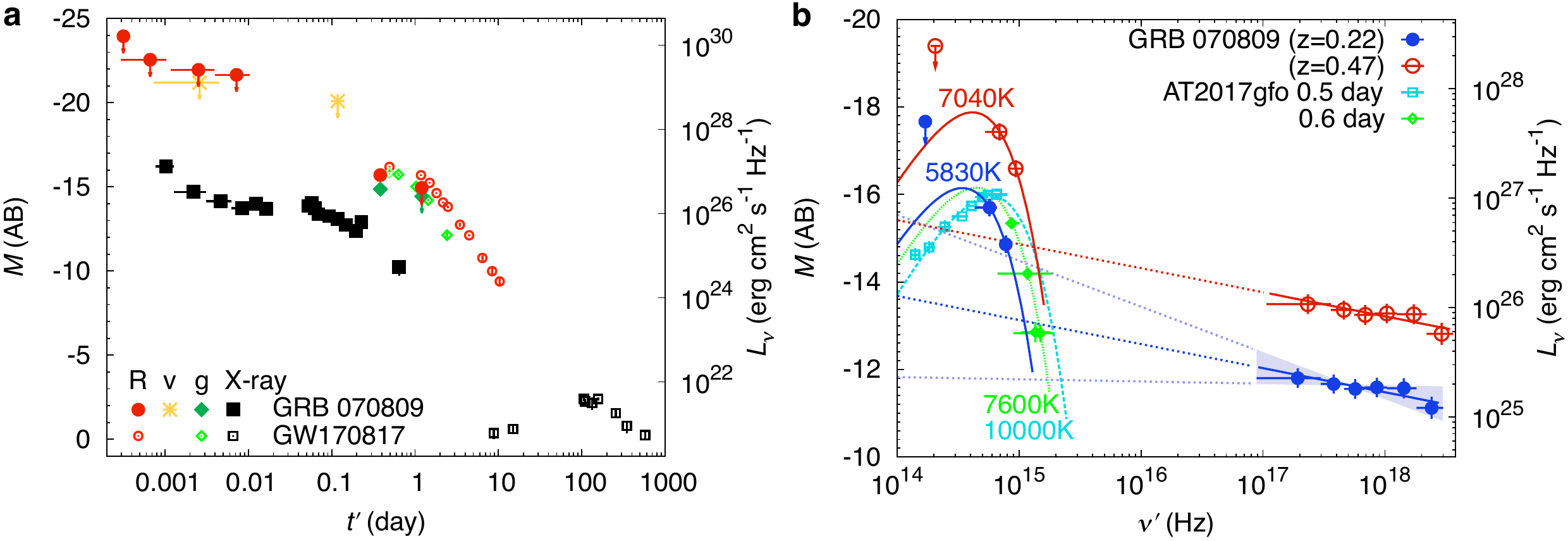}
\end{center}
\caption{
{\bf Lightcurves and SEDs of GRB 070809.}
{\bf a}, Optical $(R, g)$ and X-ray lightcurves of the afterglow of GRB 070809 (red, blue and black larger symbols) in comparison with the counterpart for GW170817 (smaller symbols), for clarity just the case of $z=0.2187$ has been shown.
{\bf b}, SEDs of the afterglow of GRB 070809 (larger symbols) in comparison with the counterpart for GW170817  (smaller symbols), and the solid (empty) circles for GRB 070809 represent the case of $z=0.2187~(0.473)$.
The optical data have been corrected for the Galactic extinction $E(B-V)=(0.08,~0.11)$ mag for GRB 070809 and GW170817/AT2017gfo, respectively.
Proper redshift corrections have been applied to GW170817 and GRB 070809.
The color term for GW170817 is derived by assuming a thermal spectrum with temperatures adopted from Ref.\cite{Drout2017}.
X-ray fluxes for both bursts are at the intrinsic energy of 1.73 keV.
The X-ray spectrum $f_{\nu}\propto \nu^{-0.22\pm0.20}$ is based on the measurements in the late $t^{-1}$ decline phase (i.e., $t>5000$ s after the trigger of the burst).
We have fitted the X-ray and optical data with a single power-law spectrum and a thermal component plus a power-law component, respectively, and got $\Delta\chi^2=12.1$. Correspondingly, the two-component model is favored over the single power-law model at a confidence level of 99.9\%, as found in our Monte Carlo simulation. The two component fits are shown in blue and red solid lines, and the shaded regions represent the $1\sigma$ uncertainties of X-ray spectra. All errors are 1$\sigma$ statistical errors and the upper limits are at the 3$\sigma$ confidence level.
}
\label{fig:GRB070809}
\end{figure}

\newpage

\begin{figure}[!h]
\begin{center}
\includegraphics[width=0.75\columnwidth]{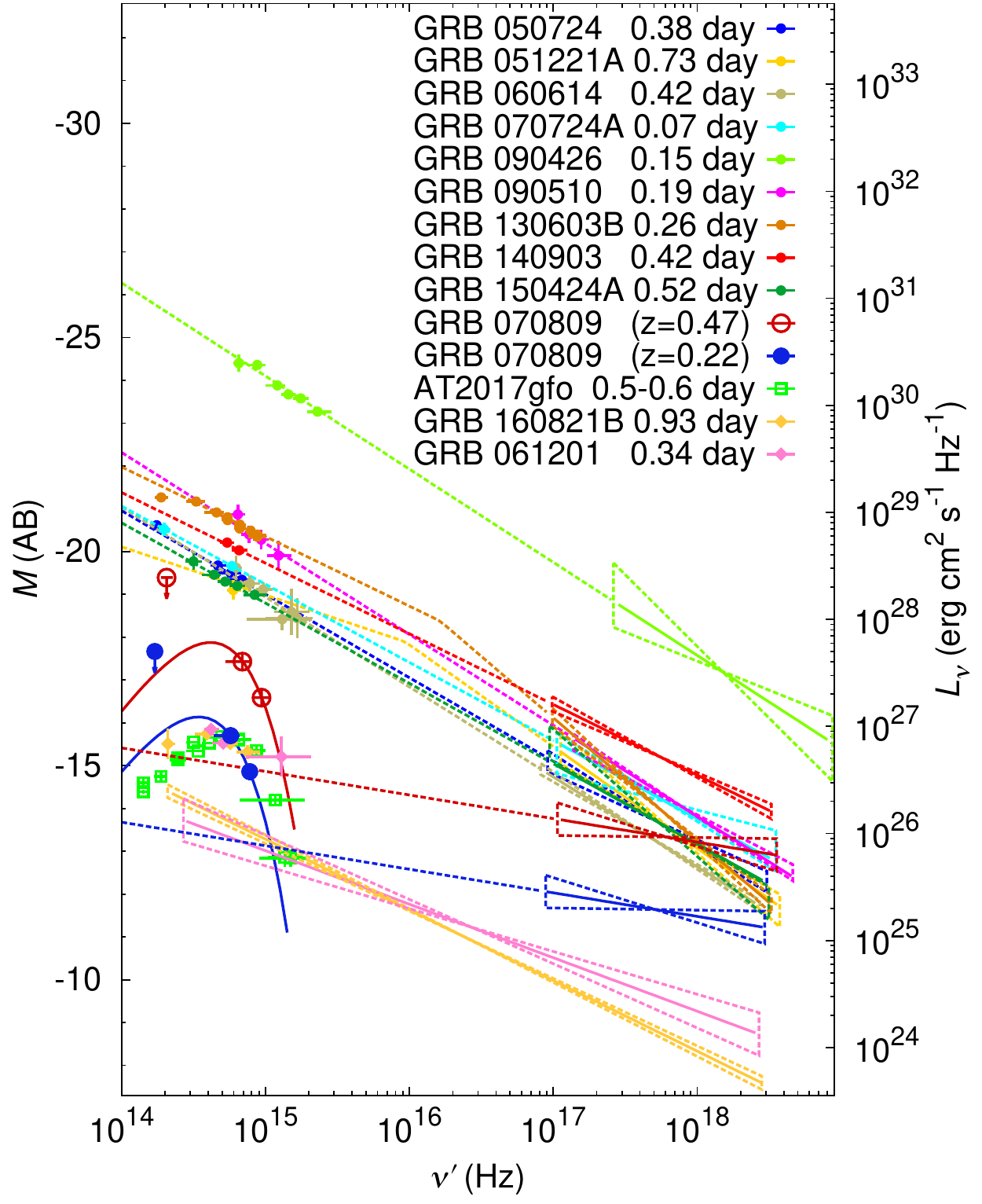}
\end{center}
\caption{
{\bf The optical to X-ray SEDs of short GRBs.} All sGRBs in this figure have bright/``plentiful" afterglow observation within 1 day after the bursts.
Symbols in different colors/shapes represent the optical data of different bursts and the redshift-corrected measurement times  are marked.
Some key properties of the events are summarized in Supplementary Table 2.
GRB 070809 is shown in larger symbols and the cases of $z=(0.2187,~0.473)$ are in blue and red, respectively.
The solid lines extended by dashed lines are the best fitted optical to X-ray spectra, and the dashed bowtie regions represent the uncertainties of X-ray spectra (while for GRB 061201 and GRB 160821B, such regions reflect the uncertainties inferred from earlier optical to X-ray fits; see Supplementary Figure 1 for the details).
All error bars represent the 1$\sigma$ statistical errors and the upper limits are at the 3$\sigma$ confidence level.
}
\label{fig:All_SED}
\end{figure}

\newpage

\begin{figure}[!h]
\begin{center}
\includegraphics[width=0.75\columnwidth]{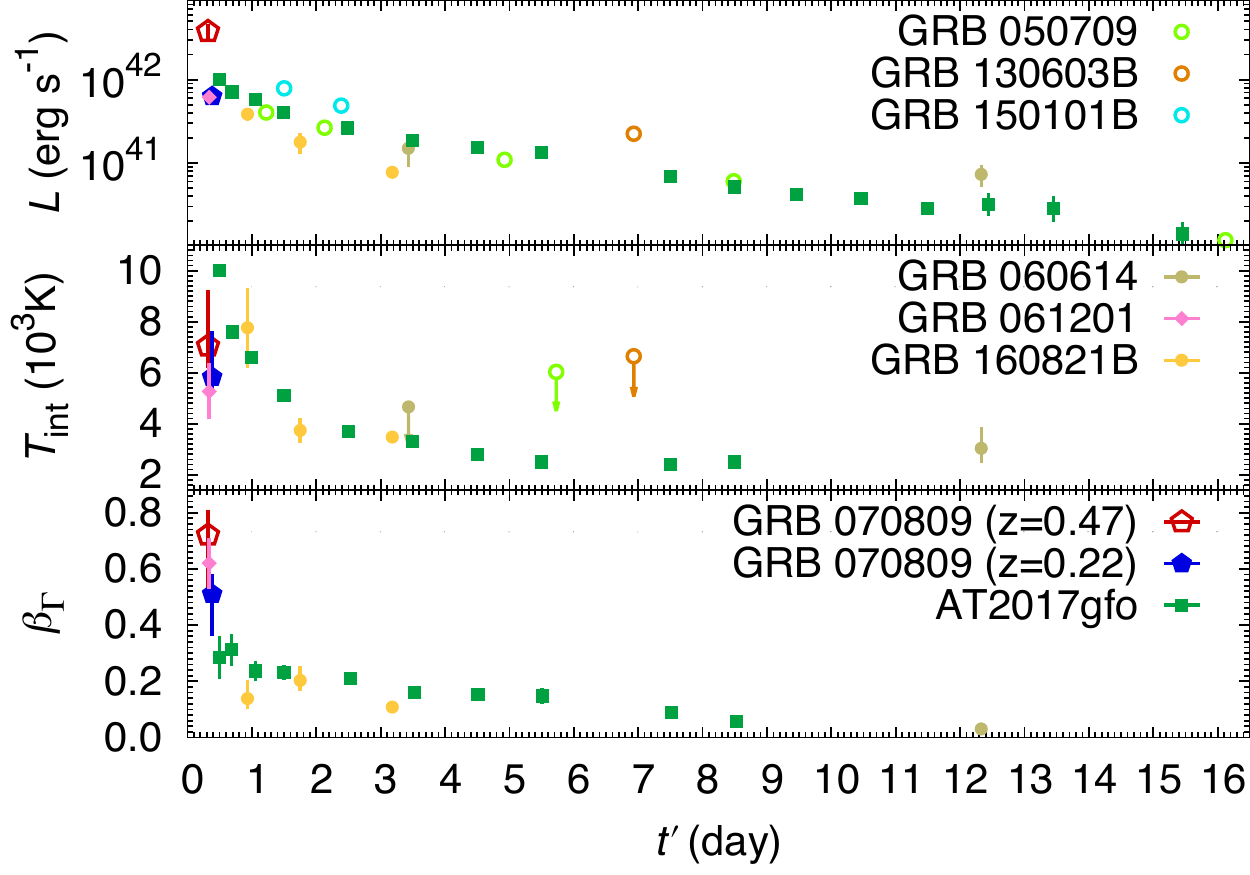}
\end{center}
\caption{
{\bf Comparison of kilonova signal of GRB 070809 with other kilonova event and candidates.}
An approximation of $L$ is adopted if a reliable temperature is unavailable, and such events are represented by empty circles without error bars. In the case of $z=0.2187$ (the blue filled pentagon), GRB 070809 has a luminosity rather similar to AT2017gfo. A redshift of 0.473 (the red empty pentagon) will instead render the kilonova signal of GRB 070809 to be the most luminous and distant one identified so far. All error bars represent the 1$\sigma$ statistical errors and the upper limits are at the 3$\sigma$ confidence level.
}
\label{fig:kilonovae}
\end{figure}

\end{addendum}

\newpage
\begin{center}
{\bf Methods}
\end{center}

{\bf The optical data analysis of GRB 070809.}
SGRB 070809 was detected by the BAT on board {\it Swift} satellite at 19:22:17 (UT) on 9 August 2007.
The BAT light curve is composed by a single short peak with a duration of $T_{90}(15-350~{\rm keV})=1.3\pm0.1$ s and
the time-averaged spectrum is best fitted by a single power-law with a photon index of $1.69\pm0.22$ (ref.\cite{Marshall2007}). {\it Swift}-UVOT began observing the field at $t\sim 74$ seconds after the BAT trigger but did not find any afterglow \cite{Marshall2007}.
The ground-based ROTSE-IIIc telescope started the observations even more promptly (i.e., 30.9 seconds after the trigger) but did not detect the afterglow either \cite{Rykoff2007}.
At about 0.47 day after the trigger, the Keck telescope observed the field and detected the optical afterglow (PI: Elizabeth Barton, Keck proposal 2007B ID:U044L, see Fig.1 for some images).
The Gemini telescope started to take images in $K$ band 40 minutes earlier but no infrared afterglow emission was detected (PI: Paul Price, Gemini proposal ID:GN-2007B-Q-27).
The next night Keck telescope revisited the field (PI: Joshua Bloom, Keck proposal 2007B ID:U138L) and the afterglow had faded away\cite{Perley2007a}.
HST observed the burst region in F606W band on 9 August 2009 and F160W band on 8 May 2010 (PI: Andrew Fruchter, HST proposal ID: 11669; see Fig.1 for one image).

For the Keck telescope observations,  we have downloaded the public archive data together with the necessary calibration frames including bias, flat and standard stars.
Standard recipes (bias subtraction, flat-field normalization and so on) were applied for data reduction.
The science frames in the same band and same night were simply added.
The optical counterpart is very close to a bright star at (13:35:03.597, -22:08:42.30) and the background is thus contaminated. The source and the standard star flux were measured using small apertures to maximize the S/N ratio and finally corrected to infinite apertures. The background, estimated in the ``blank" regions which are at angular separation from the bright star similar to that of GRB 070809, has been subtracted from the signal, the 1$\sigma$ error caused by the background variation is about 0.1 mag in both $g'$ and $R$ bands. The uncertainty introduced by the gradient of background due to the bright star is found to be 0.1 mag, too.
A standard star PG0231+051, SDSS $g'=15.729\pm0.004$ mag (AB) and Johnson $R=16.240\pm0.008$ mag (Vega, which has been converted into the AB magnitude scale with a shift of $+0.21$ mag and the induced error is about 0.02 mag),
was observed 5 times in both $g$ and $R$ bands on 11 August 2007.
These frames are used to determine zero point of that night.
In the application to the science images on 11 August 2007, the atmosphere extinctions of 0.15 mag/airmass in $g$ band and 0.11 mag/airmass in $R$ band have been taken into account
(atmosphere extinctions were taken from experimental extinction value of Mauna Kea, Hawaii, which are available at https://www.gemini.edu/sciops/telescopes-and-sites/observing-condition-constraints/extinction). The uncertainty introduced in this process is estimated to be 0.05 mag.
The total error is thus the above four uncertainties added in quadrature, which is $\approx 0.15$ mag. As a conservative approach, the $1\sigma$ errors of the optical fluxes have been finally taken to be $~0.2$ mag.
Due to the absence of standard star observation on 10 August 2007, we calibrated the science images with some bright unsaturated reference stars that were well measured on both 10 and 11 August 2007.
All these procedures are performed by means of IRAF tools (http://iraf.noao.edu).

For Gemini, the data reduction is essentially similar to Keck except one more step to subtract the dark current is applied.
There was no standard  observation in that night, and a statistical zero point 23.43 for $K$ is adopted in our analysis.
For HST data, we downloaded the full reduced production data, and measured the RMS of the GRB position. The 3$\sigma$ upper limits are then set accordingly.
The optical data analysis results are reported in Supplementary Table 1.


{\bf The X-ray data analysis of GRB 070809.} {\it Swift} XRT began to observe the field at $t\sim 71$ seconds after the BAT trigger and observed the field of GRB 070809 for three times (ObsIDs: 00287344000, 00287344001 and 00287344002). The XRT photon counting (PC) mode data are analysed by the FTOOLS software version 6.22.1 and initial event cleaning been carried out with {\tt xrtpipeline} using standard quality cuts. Then the source spectra and light curve within a circular region with a radius of 20 pixels are extracted with {\tt xselect} while the background ones are also extracted via a larger circle (i.e. 50 pixels) in a blank area. We produce the ancillary response file with {\tt xrtmkarf} to facilitate the spectral analysis, in which the response files were taken from the {\tt CALDB} database. The grouped spectra are required to have at least 1 count per bin using the {\tt cstat} approach and the parameter of absorption is set as the Galactic value (i.e. $N_{\rm H}=8.59\times 10^{20}$ $\rm cm^{-2}$) during the analysis since there is no evidence for significant absorption at the burst site \cite{Marshall2007}. A hardening of the XRT spectra is suggested by the hardness ratio data (see http://www.swift.ac.uk/xrt$_{-}$curves/00287344/). Therefore we divide the XRT data in two intervals, with an early one (from $\sim 80$ s to 1958 s) and a late one ($t>5322$ s) that follows a $\sim t^{-1}$ decline. The early and the late XRT spectra are well fitted by $f\propto \nu^{-0.60\pm 0.28}$ and $\propto \nu^{-0.22\pm0.20}$, respectively, which do show some evidence for a spectral hardening. Our results are well consistent with the  automatic analysis results provided by the {\it Swift} XRT team (http://www.swift.ac.uk/xrt$_{-}$spectra/00287344/, see ref.\cite{Evans2009} for the analysis methods), which are $\nu^{-0.57^{+0.37}_{-0.14}}$ and $\nu^{-0.21^{+0.18}_{-0.16}}$, respectively.  Meanwhile, exposure correction of the light curve has been performed by {\tt xrtlccorr} and it is binned with roughly 20 counts per bin. The exception is the last time bin (i.e., the third XRT observation) which contains in total eight net photons.

{\bf The SED sample.} The SED sample presented in Fig.3 consists of 13 events (see also Supplementary Table 2), each has a measured redshift and relatively plentiful early time (i.e., $t'<1$ day) afterglow data\cite{Malesani2007,Soderberg2006,Burrows2006,Mangano2007,Stratta2007,Berger2009,Kocevski2010,Antonelli2009,Nicuesa2012,deUgartePostigo2014,Troja2016,Knust2017,Kasliwal2017b,Coulter2017,Hjorth2017,Troja2018b,DAvanzo2018,Piro2019} that can be used to construct a reliable optical/X-ray spectrum.
The Galactic extinction corrections have been made for all bursts\cite{Schlafly2011}.

For GRB 130603B and GRB 140903A, the well measured radio to X-ray data unambiguously suggest the presence of serious dust extinction, while for GRB 070724A, the situation is less clear\cite{Berger2009}.
In any case, it is straightforward to show that the very red near-infrared/optical emission unlikely have a thermal origin (i.e., it is not a kilonova-like transient).  For a thermal origin transient, the bulk Lorentz factor of the emitting region is governed by $(1+\beta_{\Gamma})\beta_{\Gamma}\Gamma \approx 0.9L_{42}^{1/2}(T_{\rm int}/6000~{\rm K})^{-2}(1+z)(t/0.47~{\rm day})^{-1}$. For $z=0.451$ and a $K$-band flux of $\sim 9.3\mu {\rm Jy}$ at $t\sim 0.118$ day after the burst, we have $L\sim 10^{43}~{\rm erg~s^{-1}}$. With $T_{\rm int} \sim 3000$ K, we have $\Gamma \sim 50$, which is unrealistically high. We thus conclude that GRB 070724A, GRB 130603B and GRB 140903A all suffer from serious dust extinctions, which have been properly corrected \cite{Berger2009,deUgartePostigo2014,Troja2016} in Fig.3.

For GRB 061201 (GRB 160821B), the optical to X-ray SED at $t\sim 0.37$ day ($\sim 1$ day) itself is insufficient to identify an additional soft component. However, these two events have optical to X-ray SEDs constructed at times of $t\leq 0.1$ day, which are well consistent with the fireball forward shock afterglow emission model\cite{Stratta2007,Troja2019,Lamb2019} and the optical/X-ray bands are between the typical synchrotron radiation frequency ($\nu_{\rm m}$) and the cooling frequency ($\nu_{\rm c}$) of the shock-accelerated electrons. In the standard fireball afterglow model, no noticeable change of optical to X-ray SED is expected in the time interval of $0.05-1$ day, which is clearly in tension with the observation data (see  Supplementary Figure 1). The significant softening of the optical spectra in the first day after the triggers of GRB 061201 and GRB 160821B may favor the emergence of kilonova signals (For GRB 061201, the {\it Swift} UVW1 band data point seems to be well above the thermal spectrum extrapolation. However the significance of such an observation is very low\cite{Stratta2007}).
The identification of ``additional" optical
radiation component in GRB 061201 is likely benefited from the rapid ($t^{-2}$ like) decay of the forward shock emission at $t>0.7$ hour
after the trigger of the burst\cite{Stratta2007}.

{\bf Physical origins of X-ray and optical emission of GRB 070809}. The very unusual optical to X-ray spectrum in turn sheds valuable light on the underlying physics. In the standard external forward shock afterglow model, the optical spectrum can {\it not} be softer than the X-ray \cite{Piran2004}. One can {\it artificially} assume a serious extinction in the ``host" galaxy to yield an optical spectrum as hard as that of the X-rays, the corresponding X-ray flux however would be significantly in excess of the observed. Hence it is fairly robust to conclude that the optical and X-ray emission of GRB 070809 have different physical origins.
There are two possibilities:
(i) the optical emission was the forward shock emission while the X-ray afterglow was  powered by the prolonged activity of the central engine; (ii) the optical afterglow was the blue kilonova emission while X-ray afterglow is attributable to either the prolonged activity of the central engine or the forward shock emission.  Below we examine these two possibilities case by case. Since our main concern is the physical origin of the optical emission, here we only present a brief discussion of the X-ray emission. If the forward shock gave rise to the X-ray afterglow, the flat segment calls for either the energy injection from the central engine or a structured outflow, like the modelling of the X-ray flat segment of sGRB 051221A, since the standard fireball afterglow model predicts a much quicker decline (see review\cite{Berger2014} and the references therein).
An additional requirement is a rather hard shock-accelerated electron spectrum (that is, $p\sim 1.5$) to match the X-ray data. If instead the X-ray afterglow was attributed to the prompt energy dissipation of the continual outflow launched via the prolonged activity of the central engine, plausibly favored by the X-ray spectral hardening shown in the XRT data, the hard spectrum of X-ray emission may be accounted for by the fast-cooling electrons suffering from dominant inverse Compton cooling in the Klien$-$Nishina regime. The magnetic field in the emitting region, however, should be very
low otherwise the Klein$-$Nishina effect is too weak to modify the energy distribution of electrons and then the radiation spectrum\cite{Nakar2009}. While for a sGRB that is expected to be powered by the merger of a neutron star binary, a long-lasting (i.e., $10^{4}-10^{5}$ s) relativistic outflow could be launched by magnetic processes, rather than the neutrino process, given the expected very-low rate ($<10^{-4}M_\odot~{\rm s^{-1}}$) of the fall-back accretion onto the central remnant\cite{Fan2005} that can only yield a relatively cool accretion disk. Therefore, in the central engine activity model for the X-ray emission, a hard spectrum of electrons accelerated by the magnetic energy dissipation is still needed unless the magnetic energy dissipation was extremely efficient in the outflow acceleration phase. With a very hard spectrum of X-rays, the extrapolated optical emission will be significantly lower than the observed fluxes and hence the optical emission, dominated by a component with a different physical origin, will not be effectively influenced. This conclusion is independent of the physical progress powering the X-rays as long as these photons originated from the synchrotron radiation.

Below we focus on the more intriguing optical emission. As for the forward shock emission scenario, the main challenge is the unusually red optical spectrum. With the introduction of a high extinction of the ``host" galaxy, for example $A_{\rm V}\sim 1$ mag for a Small Magellanic Cloud extinction light curve at $z=0.2187$,
the ``intrinsic" spectrum could become as hard as $\sim \nu^{-0.7}$,
which is still much softer than the observed X-ray spectrum $f_{\nu_{\rm X}} \propto \nu^{-0.22\pm 0.20}$. In fact, the very low intrinsic hydrogen column
density (i.e., $N_{\rm H}=0^{+3.42}_{-0}\times10^{20}{\rm cm}^{-2}$, as inferred in X-ray spectrum modeling) indicates a negligible extinction. Moreover, at a redshift of $z=0.2187$ or $0.473$, the absence of a host galaxy down to $28$th mag can only be met by a faint dwarf galaxy like those belonging to the Local Group or the burst took place outside of the host galaxy, for which no strong extinction is expected (note that currently all short/long-short bursts with reliable measurement of a strong extinction have been identified within or on the outskirts of the bright host galaxies, as summarized in Supplementary Table 2). Therefore, there is no motivation to assume strong extinction in the host galaxy. The interpretation of the very soft optical spectrum would thus require an extremely large $p\sim 5$, which in turn would yield a decline of
$\propto t^{-3}$, too quick to be consistent with the current data (i.e., the decline should be shallower than $t^{-0.96}$). We conclude that the first scenario is strongly disfavored.

The second scenario seems to be much more natural. If interpreted as a kilonova component at $z=(0.2187,0.473)$ without suffering from significant extinction, the intrinsic temperature is $T_{\rm int}\approx (5830^{+1800}_{-700},~7040^{+2170}_{-840})$ K, which seems lower than that of AT2017gfo at a similar time (see Fig.4). The isotropic-equivalent luminosity of a relativistic outflow with a bulk Lorentz factor $\Gamma$ (the corresponding velocity, in unit of the speed of light ($c$), is denoted as $\beta_{\Gamma}$) can be described as $L \approx 4\pi R^2 \sigma_{\rm T}T_{\rm int}^4\Gamma^{-2}$, where $\sigma_{\rm T}$ is the Stefan$-$Boltzmann constant, and $R\approx \beta_{\rm \Gamma}ct/[(1-\beta_{\Gamma})(1+z)]$ is the radius of the emitting region. Hence we have $(1+\beta_{\Gamma})\beta_{\Gamma}\Gamma \approx 0.9L_{42}^{1/2}(T_{\rm int}/6000~{\rm K})^{-2}(1+z)(t/0.47~{\rm day})^{-1}$. For $z=(0.2187,~0.473)$, we have $L\approx (0.63,3.8)\times 10^{42}~{\rm erg~s^{-1}}$ and hence $\beta_{\Gamma}\approx (0.51^{+0.07}_{-0.15},~0.72^{+0.09}_{-0.2})$.
The mass of the wind is estimated as $M_{\rm ej}\sim 0.018~{M_\odot}~(\beta_{\Gamma}/0.5)(t/0.5~{\rm day})^{2}[(1+z)/1.22]^{-2}(\kappa/0.2~{\rm cm^2~g^{-1}})^{-1}$ (for $z=0.473$ and $\beta_\Gamma \sim 0.72$ we  have $M_{\rm ej}\sim 0.016M_\odot$), which is comparable to that inferred from the modeling of AT2017gfo, where $\kappa$ is the opacity parameter of the sub-relativistic lanthanide-poor/free outflow.  Though the initial dynamical ejecta of compact object mergers are expected to be lanthanide-rich, due to the absorption of neutrinos from the nascent precollapse massive neutron star or the accretion disk at high latitudes, some material will become lanthanide-poor/free. A relatively long-lived accretion disk can also launch lanthanide-poor/free material. Our finding is support of these theoretical predictions\cite{Metzger2014,Kasen2015,Metzger2017}. The red kilonova component of GRB 070809, which should have lasted 1-2 weeks, however was too dim ($\geq 26$th AB magnitude in H-band at $t\sim 10$ day after the burst if we simply shift AT2017gfo to $z\geq 0.2187$) to be caught even by ground-based very large size telescopes at such a redshift. The above blue/red kilonova scenarios are the most widely adopted ones for similar phenomena in the literature.
Certainly, some other interesting possibilities, such as the shock breakout/cocoon models\cite{Kasliwal2017}  and the models with energy injection from the central engine (see a comprehensive review\cite{Metzger2017} for the details), can not be ruled out. Particularly, the shock breakout/cocoon model may better account for the transrelativistic velocity $\beta_\Gamma \sim 0.72$ in the case of $z=0.473$.

{\bf The kilonova sample.}  Before 2017, evidence for kilonova emission had been identified in sGRB 130603B at $z=0.356$ based on a remarkable excess of the F160W-band flux by a large factor of $\sim 50$  \cite{Tanvir2013,Berger2013}, GRB 060614 at $z=0.125$ thanks to the late time optical spectral softening\cite{Yang2015,Jin2015} and sGRB 050709 at $z=0.16$ because of the different decline rates in $I/F814W$ and $R$ bands\cite{Jin2016}. In the absence of gravitational wave detection, such signals had been widely taken as the most convincing evidence for the compact object merger origin of some GRBs. After the successful detection of AT2017gfo\cite{Pian2017,Drout2017,Kasliwal2017,Arcavi2017,Smartt2017,Cowperthwaite2017,Chornock2017,Valenti2017}, a few kilonova candidates were reported in the literature. For instance, the $R$ band emission of GRB 150101B detected at $t>1$ day after its trigger\cite{Fong2016} was found to be brighter than that predicted by the radio-X-ray spectrum and might consist of a blue kilonova component\cite{Troja2018} that resembles the signal of GRB 050709\cite{Jin2016}. Another kilonova signal, initially identified at $t\sim 3.6$ days after the trigger of GRB 160821B\cite{Jin2018}, has been confirmed with substantially extended data sets\cite{Troja2019,Lamb2019} during the revision of this work.

Among kilonova events/candidates, AT2017gfo has the most plentiful data, with which the evolution of the luminosity, the temperature and the velocity of the emitting region have been reliably inferred in the time range of $\sim 0.5-13.5$ days.  This provides a benchmark for comparison with other new events/candidates.
For GRB 060614, the temperature was only reliably measured at $t\sim 13.6$ days\cite{Jin2015}.  The temperature of the kilonova signal of GRB 160821B was firstly suggested to be $\sim 3100$ K at $t\sim 3.6$ days after the burst\cite{Jin2018}. With the substantially extended data set reported recently\cite{Troja2019,Lamb2019}, the evolutions of the $T_{\rm int}$, $L$ and $\beta_\Gamma$ can be constructed (Fig.4; the details will be presented in Li et al. 2019, in preparation).
Upper limits on $T_{\rm int}$ for GRB 130603B and GRB 050709 can be set.
For GRB 150101B, no optical SED can be constructed due to the sole detection in R-band. Therefore, the bolometric luminosities are unavailable for GRB 130603B, GRB 050709 and GRB 150101B and we simply adopt the approximation of $1.4\nu L_\nu$ (for a thermal spectrum we have $L\approx 1.4\nu_{\rm p} L_{\nu_{\rm p}}$, where $L_\nu=4\pi D_{\rm L}^{2}f_\nu$ and $\nu_{\rm p}$ is the frequency at which the energy flux $\nu f_\nu$ peaks). That is why in Fig.4 these three events are marked in open circles without error bars. The absence of reliable $T_{\rm int}$ for GRB 130603B, GRB 050709 and GRB 150101B also hampers the estimate of $\beta_\Gamma$.

{\bf Code availability.} The codes used in this analysis are standard in the community, as introduced in the Methods.

{\bf Data availability.} The Keck, HST, Gemini and {\it Swift} observation data analysed/used in
this work are all publicly available.

\renewcommand{\refname}{References}


\newpage

\begin{center}
{\bf Supplementary Information}
\end{center}

\begin{table}[!ht]
\label{tab:GRB070809}
\begin{center}
\title{}{\bf Supplementary Table 1. Observations of GRB070809. }\\
\begin{tabular}{lllll}\hline\hline
$t$	& Exposure	& Instrument	& Filter	& Magnitude$^{a}$\\
(days)	& (seconds)	& 	& 	& (AB) \\ \hline
0.44348	& 900	& Gemini-N+NIRI	& K  & ($>22.4$) \\
0.46746	& 1320	& Keck I+LRIS	& g  & $25.48\pm0.20$ \\
0.46767	& 1200	& Keck I+LRIS	& R  & $24.54\pm0.20$ \\
1.47001	& 820	& Keck I+LRIS	& g  & ($>25.9$) \\
1.46867	& 580	& Keck I+LRIS	& R  & ($>25.3$) \\
730.94	& 5150	& HST+ACS	& F606W  & ($>28.0$) \\
1002.80	& 5597	& HST+WFC3	& F160W  & ($>26.2$) \\
\hline\hline
\end{tabular}
\begin{tabular}{l}
Note: a. These values have not been corrected for the Galactic extinction \\
of $E(B-V)=0.08$ mag (ref.\cite{sup_Schlafly2011}). All errors represent the 1$\sigma$ statistical \\
errors and the upper limits are at the 3$\sigma$ confidence level. \\
\end{tabular}
\end {center}
\end{table}

\renewcommand{\figurename}{Supplementary Figure}
\setcounter{figure}{0}
\begin{figure}[!ht]
\includegraphics[width=0.49\columnwidth]{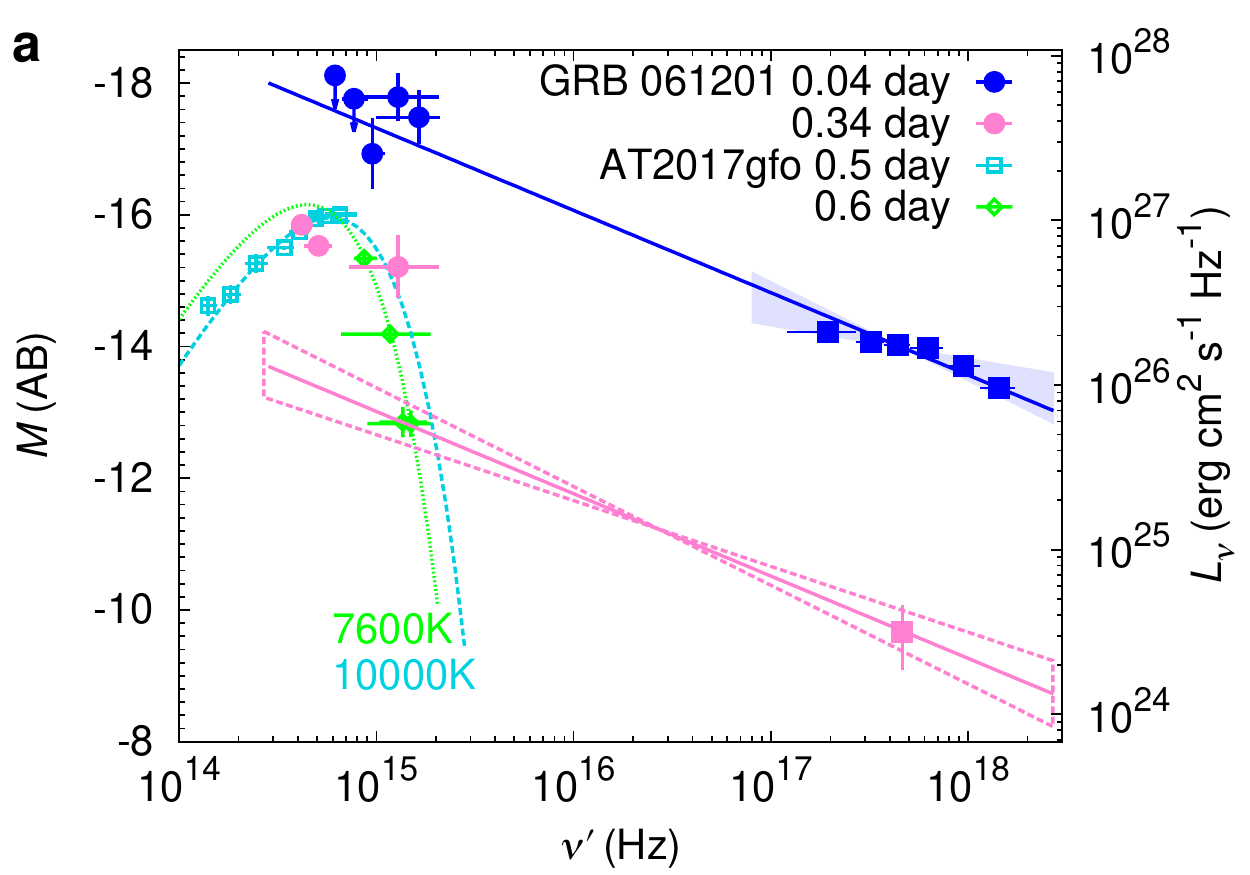}
\includegraphics[width=0.49\columnwidth]{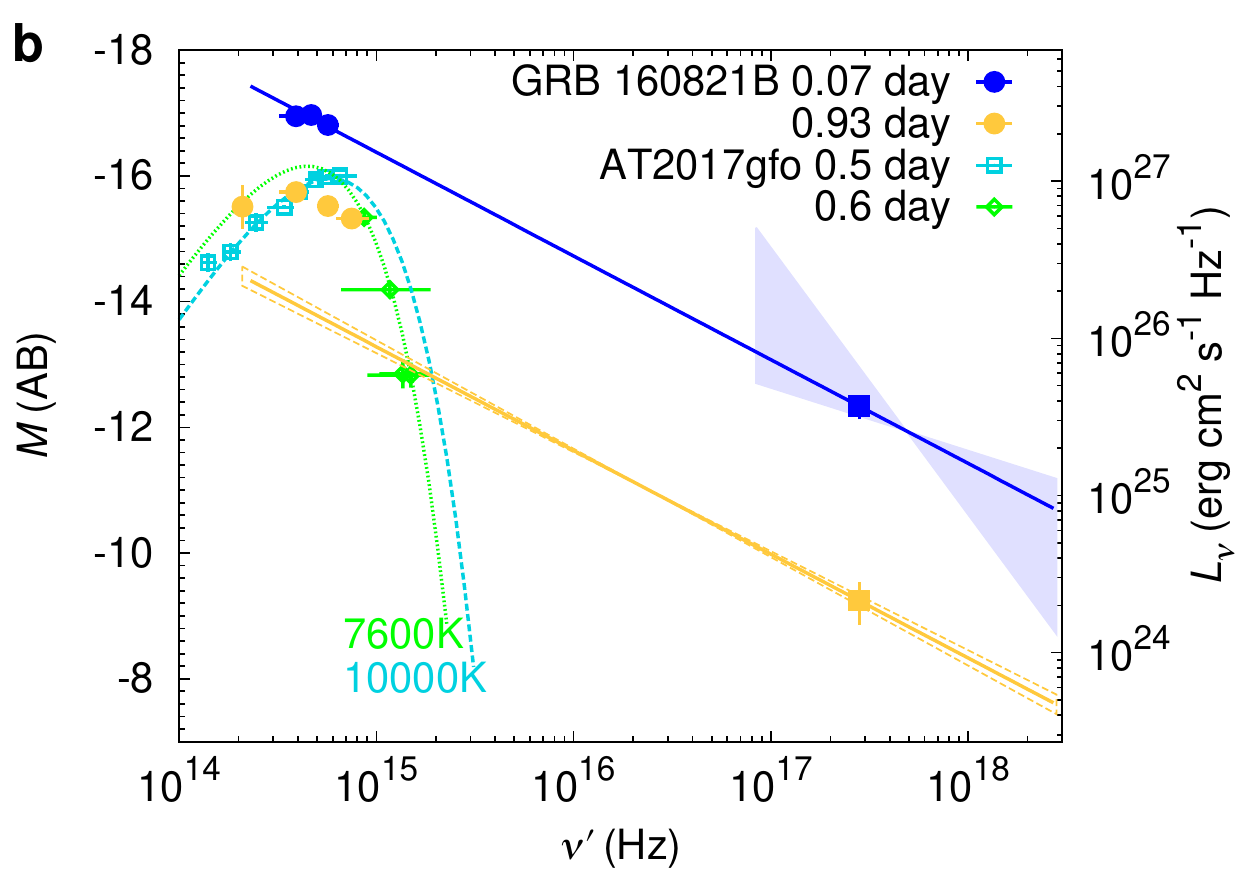}
\caption{
{\bf The optical to X-ray SEDs of GRB 061201 (a) and GRB 160821B (b).} The optical and X-ray data are marked by filled circles and squares, respectively.
At earlier times (i.e. $t'\sim 0.04~{\rm day}$ for GRB 061201 and $t'\sim 0.07~{\rm day}$ for GRB 160821B), the optical to X-ray SEDs can be reasonably fitted by single power-laws, and the power-law indexes are well consistent with the forward shock afterglow models \cite{sup_Stratta2007,sup_Troja2019,sup_Lamb2019}, where $t'$ represents the time measured in the source frame. The best fit SEDs are represented by the blue solid lines and the blue bowtie shades define the expected $1\sigma$ region of the X-ray fluxes.  While at late times (i.e., $t'\sim 0.34$ day for GRB 061201 and $t'\sim 0.93$ day for GRB 160821B), the forward shock afterglow spectra inferred at earlier times (i.e., $f_\nu \propto (\nu^{-0.5\pm0.1}, \nu^{-0.66\pm 0.03})$ for these two GRBs \cite{sup_Stratta2007,sup_Lamb2019}, which are in pink and yellow lines, respectively), can not fully account for the data. Instead there are optical emission excesses which may be attributed to thermal-like emission with $T_{\rm int}\sim 4741\pm987$K and $T_{\rm int} \sim 6689\pm1350$K, respectively. Kilonova emission from AT2017gfo at a similar time is also plotted for comparison, see the green open symbols.
All error bars represent the 1$\sigma$ statistical errors and the upper limits are at the 3$\sigma$ confidence level.
}
\label{fig:GRB061201}
\end{figure}
\vfill

\begin{table}[!ht]
\begin{center}
\label{tab:comparison}
\title{}{\bf Supplementary Table 2. Comparison of the properties of some sGRBs.} \\
\begin{tabular}{lllllllll}\hline\hline
Event	& z	& A$_{\rm V}~^{\rm a}$	& A$_{\rm V}~^{\rm a}$	& $\beta_{\rm oX}~^{\rm b}$	& Galactic N$_{\rm H}$	& Host N$_{\rm H}$ & In Host	& Ref.	\\
		& 	& Galactic		& Host		&	& ($\times10^{20}$)	& ($\times10^{20}$)	& 	&	\\
\hline
GRB 050724	& 0.257	& 1.43	& 0				& $-0.78$	& 27.7	& $0^{+10.3}_{-0}$	& Yes	& (1) \\
GRB 051221A	& 0.5465	& 0.18	& 0				& $-0.45({\rm B})^{\rm c}$	& 7.52	& $9\pm6$	& Yes	& (2-3) \\
GRB 060614	& 0.125	& 0.07	& 0.05			& $-0.84$	& 1.99	& $1.7^{+2.4}_{-1.7}$	& Yes	& (4) \\
GRB 061201	& 0.111	& 0.20	& 0				& $-0.50$	& 6.57	& $6^{+7}_{-6}$	& Hostless	& (5) \\
GRB 070724	& 0.457	& 0.12	& $1.6^{\rm d}$	& $-0.73$	& 1.21	& $2.5^{+32}_{-2.5}$	& Yes	& (6-7) \\
GRB 070809	& 0.2187/0.473	& 0.25	& 0				& $-0.22$	& 8.59	& $0^{+3.42}_{-0}$	& Hostless 	& (8) 	\\
GRB 090426	& 2.609	& 0.04	& 0				&$-0.87$	& 1.58	& $23^{+56}_{-19}$	& Yes	& (9) \\
GRB 090510	& 0.903	& 0.06	& 0.17			&$-0.85$	& 1.77	& $34^{+38}_{-30}$	& Yes	& (10)\\
GRB 130603B	& 0.356	& 0.06	& 0.9	& $-0.65({\rm B})^{\rm c}$ & 2.10	& $43.8^{+10}_{-9.1}$	& Yes	& (11) \\
GRB 140903	& 0.351	& 0.08	& 0.4	& $-0.66$ & 3.26	& $18^{+9}_{-8}$	& Yes	& (12) \\
GRB 150424A	& 0.3	& 0.16	& 0				& $-0.75$	& 6.02	& $9^{+6}_{-5}$	& Hostless	& (13) \\
GRB 160821B	& 0.16	& 0.12	& 0				& $-0.70$	& 5.75	& $0^{+3.45}_{-0}$	& Hostless	& (14) \\
GW170817	& 0.0098	& 0.33	& 0				& no fit	& 8.76	& 0	& Yes	& (15-19) \\
\hline\hline
\end{tabular}
\begin{tabular}{l}
(1). Malesani et al. 2007\cite{sup_Malesani2007}; (2). Soderberg et al. 2006\cite{sup_Soderberg2006}; (3). Burrows et al. 2006\cite{sup_Burrows2006}; (4). Mangano et al. 2007\cite{sup_Mangano2007}; \\
(5). Stratta et al. 2007\cite{sup_Stratta2007}; (6). Berger et al. 2009\cite{sup_Berger2009}; (7). Kocevski et al. 2010\cite{sup_Kocevski2010}; (8). This work; (9). Antonelli et al. 2009\cite{sup_Antonelli2009}; \\
(10). Nicuesa Guelbenzu et al. 2012\cite{sup_Nicuesa2012}; (11). de Ugarte Postigo et al. 2014\cite{sup_deUgartePostigo2014}; (12). Troja et al. 2016\cite{sup_Troja2016}; \\
(13). Knust et al. 2017 \cite{sup_Knust2017}; (14). Kasliwal et al. 2017b\cite{sup_Kasliwal2017b}; (15). Coulter et al. 2017\cite{sup_Coulter2017}; (16). Hjorth et al. 2017\cite{sup_Hjorth2017}; \\
(17). Troja et al. 2018b\cite{sup_Troja2018b}; (18). D'Avanzo et al. 2018\cite{sup_DAvanzo2018}; (19). Piro et al. 2019\cite{sup_Piro2019}.\\
a. In units of mag. \\
b. The optical to X-ray spectral index. \\
c. A broken power-law model with an index increase by a factor of 0.5 is needed.\\
d. Based on the assumption of an optical to X-ray spectral index $\beta_{\rm oX}=-0.73$.
\end{tabular}
\end{center}
\end{table}

\renewcommand{\refname}{Supplementary References}

\end{document}